\documentclass{article}
\usepackage{spconf}
\usepackage[noadjust]{cite}
\usepackage{bbm}
\usepackage{array}
\usepackage{dcolumn}
\usepackage{epsfig}
\usepackage[intlimits]{amsmath}
\usepackage{amsmath, amsfonts, yhmath, bm}
\usepackage{amssymb}
\usepackage{psfrag}
\usepackage{color,soul}
\usepackage[dvipsnames]{xcolor}
\usepackage[normalem]{ulem}
\usepackage{enumerate}
\usepackage{stackengine}
\usepackage{graphicx}
\usepackage{subcaption}
\usepackage{multirow}
\usepackage{etoolbox}
\usepackage{tcolorbox}
\usepackage{float}
\usepackage{dsfont}
\usepackage{algpseudocode}
\usepackage{tikz}
\usepackage{mathtools}
\usepackage{arydshln}
\usetikzlibrary{shapes.geometric, arrows}

\usepackage[ruled,vlined]{algorithm2e}

\usepackage{colortbl}
\usepackage{booktabs}

\usepackage{svg}
\usepackage{lipsum}
\usepackage{pgfplots}
\pgfplotsset{compat=1.18}
\usepackage{pgfplotstable}

\usepackage{hyperref}
\hypersetup{
    colorlinks,
    linkcolor={red!50!black},
    citecolor={blue!50!black},
    urlcolor={blue!80!black}
}

\usepackage{balance}
\usepackage{url}
\usepackage[inline]{enumitem} 

\usetikzlibrary{backgrounds, positioning, fit}

\usepackage{vmr-symbols-vecbold}
\usepackage{standard-macros}
\PassOptionsToPackage{hyphens}{url}
\usepackage{hyperref}

\DeclareSymbolFontAlphabet{\amsmathbb}{AMSb}%

\newcommand{\lro}[1]{\lefto({#1}\right)}																

\newcommand{\lr}[1]{\left({#1}\right)}																
\newcommand{\lrb}[1]{\left \lbrace {#1} \right \rbrace}															

\safemath{\dopplerspread}{B_D}																								
\safemath{\delayspread}{T_D}																									
\safemath{\nc}{n\sub{c}}																										
\safemath{\nf}{n\sub{f}}																										
\safemath{\efa}{p\sub{sc}}
\safemath{\efb}{p\sub{cs}}
\safemath{\ef}{\epsilon\sub{f}	}
\safemath{\nd}{n\sub{d}}																										
\safemath{\ntx}{n\sub{t}} 																											
\safemath{\nrx}{n\sub{r}}																											
\safemath{\ntxt}{\tilde{n\sub{t}}}																											
\safemath{\cb}{\ensuremath{L}} 																								
\safemath{\cl}{\ensuremath{n}} 																								
\safemath{\txanto}{{\ensuremath{\tilde{m}_t}}} 																		
\safemath{\cs}{M} 																														
\safemath{\idPustm}{\ensuremath{S_{k}}}
\safemath{\error}{\ensuremath{\epsilon}} 																				
\safemath{\eexp}{\ensuremath{\mathcal{E}}} 																			
\safemath{\nsubc}{n\sub{s}}			 																						
\safemath{\nofdm}{n\sub{o}} 																									
\safemath{\bc}{\ensuremath{B_c}} 																							
\safemath{\ts}{\ensuremath{T_s}} 																							
\safemath{\nrb}{\ensuremath{n_{rb}}} 																						
\safemath{\rul}{\ensuremath{\rho\sub{ul}}}
\safemath{\rdl}{\ensuremath{\rho\sub{dl}}}

\safemath{\nres}{\ell}
\safemath{\nr}{n\sub{r}}
\safemath{\maxk}{M^*\lr{\nres, \nsubc, \nofdm, \epsilon, \rho}}
\safemath{\Rmax}{R^*}
\safemath{\Emin}{E\sub{b}^*/N_0}
\safemath{\Eminf}{\frac{E\sub{b}^*}{N_0}}
\safemath{\np}{\ensuremath{n\sub{p}}}
\safemath{\ndf}{\ensuremath{\bar{n}\sub{d}}}
\safemath{\npf}{\ensuremath{\bar{n}\sub{p}}}
\safemath{\code}{\ensuremath{\mathcal{C}}}
\safemath{\err}{\ensuremath{\epsilon}}
\safemath{\rp}{\ensuremath{\rho\sub{p}}}
\safemath{\rd}{\ensuremath{\rho\sub{d}}}
\safemath{\cohtime}{\ensuremath{T\sub{c}}}
\safemath{\cohbw}{\ensuremath{B\sub{c}}}
\safemath{\nmax}{\ensuremath{\ell\sub{m}}}
\safemath{\ntot}{\ensuremath{n\sub{tot}}}
\safemath{\nul}{\ensuremath{n\sub{ul}}}
\safemath{\ndl}{\ensuremath{n\sub{dl}}}

\safemath{\yp}{\ensuremath{\randvecy_{\nu}^{(\text{p})}}}
\safemath{\yd}{\ensuremath{\randvecy_{\nu}^{(\text{d})}}}
\safemath{\ypd}{\ensuremath{\vecy_{\nu}^{(\text{p})}}}
\safemath{\ydd}{\ensuremath{\vecy_{\nu}^{(\text{d})}}}

\safemath{\ypf}{\ensuremath{\bar{\randvecy}_{\nu}^{(\text{p})}}}
\safemath{\ydf}{\ensuremath{\bar{\randvecy}_{\nu}^{(\text{d})}}}
\safemath{\ypdf}{\ensuremath{\bar{\vecy}_{\nu}^{(\text{p})}}}
\safemath{\yddf}{\ensuremath{\bar{\vecy}_{\nu}^{(\text{d})}}}

\safemath{\xp}{\ensuremath{\vecx^{(\text{p})}}}
\safemath{\xd}{\ensuremath{\randvecx_{\nu}^{(\text{d})}}}
\safemath{\xdd}{\ensuremath{\vecx_{\nu}^{(\text{d})}}}

\safemath{\xpf}{\ensuremath{\bar{\vecx}^{(\text{p})}}}
\safemath{\xdf}{\ensuremath{\bar{\randvecx}_{\nu}^{(\text{d})}}}
\safemath{\xddf}{\ensuremath{\bar{\vecx}_{\nu}^{(\text{d})}}}

\safemath{\xdb}{\ensuremath{\overline{\randvecx}^{(\text{d})}}}
\safemath{\Pxd}{\ensuremath{P_{\randvecx^{(\text{d})}}}}

\safemath{\xpbar}{\ensuremath{\overline{\matX}^{(\text{p})}}}
\safemath{\xdbar}{\ensuremath{\overline{\randmatX}^{(\text{d})}}}

\safemath{\xdv}{\ensuremath{\randvecx^{(\text{d})}}}
\safemath{\xdbarv}{\ensuremath{\overline{\randvecx}^{(\text{d})}}}
\safemath{\ydv}{\ensuremath{\randvecy^{(\text{d})}}}

\safemath{\xdr}{\ensuremath{\matX^{(\text{d})}}}

\safemath{\ttx}{\ensuremath{\tau\sub{tx}}}
\safemath{\trx}{\ensuremath{\tau\sub{rx}}}
\safemath{\ack}{\ensuremath{\mathrm{s}}}
\safemath{\nack}{\ensuremath{\mathrm{c}}}

\newcommand{\prob}[1]{\ensuremath{\mathbb{P}\lro{#1}}}

\safemath{\mI}{\ensuremath{i\lro{\randvecy ; \randvecx}}} 				

\safemath{\randveca}{\bm{A}}
\safemath{\randvecb}{\bm{B}}
\safemath{\randvecc}{\bm{C}}
\safemath{\randvecd}{\bm{D}}
\safemath{\randvece}{\bm{E}}
\safemath{\randvecf}{\bm{F}}
\safemath{\randvecg}{\bm{G}}
\safemath{\randvech}{\bm{H}}
\safemath{\randveci}{\bm{I}}
\safemath{\randvecj}{\bm{J}}
\safemath{\randveck}{\bm{K}}
\safemath{\randvecl}{\bm{L}}
\safemath{\randvecm}{\bm{M}}
\safemath{\randvecn}{\bm{N}}
\safemath{\randveco}{\bm{O}}
\safemath{\randvecp}{\bm{P}}
\safemath{\randvecq}{\bm{Q}}
\safemath{\randvecr}{\bm{R}}
\safemath{\randvecs}{\bm{S}}
\safemath{\randvect}{\bm{T}}
\safemath{\randvecu}{\bm{U}}
\safemath{\randvecv}{\bm{V}}
\safemath{\randvecw}{\bm{W}}
\safemath{\randvecx}{\bm{X}}
\safemath{\randvecy}{\bm{Y}}
\safemath{\randvecz}{\bm{Z}}
\safemath{\randvecphi}{\bm{\Phi}}

\safemath{\randmatA}{\amsmathbb{A}}
\safemath{\randmatB}{\amsmathbb{B}}
\safemath{\randmatC}{\amsmathbb{C}}
\safemath{\randmatD}{\amsmathbb{D}}
\safemath{\randmatE}{\amsmathbb{E}}
\safemath{\randmatF}{\amsmathbb{F}}
\safemath{\randmatG}{\amsmathbb{G}}
\safemath{\randmatH}{\amsmathbb{H}}
\safemath{\randmatI}{\amsmathbb{I}}
\safemath{\randmatJ}{\amsmathbb{J}}
\safemath{\randmatK}{\amsmathbb{K}}
\safemath{\randmatL}{\amsmathbb{L}}
\safemath{\randmatM}{\amsmathbb{M}}
\safemath{\randmatN}{\amsmathbb{N}}
\safemath{\randmatO}{\amsmathbb{O}}
\safemath{\randmatP}{\amsmathbb{P}}
\safemath{\randmatQ}{\amsmathbb{Q}}
\safemath{\randmatR}{\amsmathbb{R}}
\safemath{\randmatS}{\amsmathbb{S}}
\safemath{\randmatT}{\amsmathbb{T}}
\safemath{\randmatU}{\amsmathbb{U}}
\safemath{\randmatV}{\amsmathbb{V}}
\safemath{\randmatW}{\amsmathbb{W}}
\safemath{\randmatX}{\amsmathbb{X}}
\safemath{\randmatY}{\amsmathbb{Y}}
\safemath{\randmatZ}{\amsmathbb{Z}}
\safemath{\randmatSigma}{\mathbb{\Sigma}}
\safemath{\randmatPhi}{\mathbb{\Phi}}
\safemath{\randmatLambda}{\mathbb{\Lambda}}

\safemath{\matSigma}{\bm{\Sigma}}
\safemath{\matPhi}{\bm{\Phi}}
\safemath{\matLambda}{\bm{\Lambda}}

\usepackage{glossaries}
\glsdisablehyper
\newglossaryentry{mie}
{
  name={MIE},
  description={maximum index-based estimator},
  first={\glsentrydesc{mie} (\glsentrytext{mie})}
}
\usepackage{pgfplots}
\pgfplotsset{compat=1.14}
\usepgfplotslibrary{groupplots}
\usetikzlibrary{pgfplots.groupplots}

\definecolor{mygray}{gray}{0.9}

\makeatletter
\newcommand\footnoteref[1]{\protected@xdef\@thefnmark{\ref{#1}}\@footnotemark}
\makeatother

\usepackage[lining]{ebgaramond}

\makeatletter
\newsavebox\myboxA
\newsavebox\myboxB
\newlength\mylenA

\newcommand*\mybar[2][0.75]{%
	\sbox{\myboxA}{$\m@th#2$}%
	\setbox\myboxB\null
	\ht\myboxB=\ht\myboxA%
	\dp\myboxB=\dp\myboxA%
	\wd\myboxB=#1\wd\myboxA
	\sbox\myboxB{$\m@th\overline{\copy\myboxB}$}
	\setlength\mylenA{\the\wd\myboxA}
	\addtolength\mylenA{-\the\wd\myboxB}%
	\ifdim\wd\myboxB<\wd\myboxA%
	\rlap{\hskip 0.5\mylenA\usebox\myboxB}{\usebox\myboxA}%
	\else
	\hskip -0.5\mylenA\rlap{\usebox\myboxA}{\hskip 0.5\mylenA\usebox\myboxB}%
	\fi}
\makeatother

\makeatletter
\newcommand\customsize{\@setfontsize\customsize{11}{13.6}}
\makeatother

\title{Achieving Robustness in Blind Modulo Analog-to-Digital Conversion}

\name{Amir Weiss}

\address{Faculty of Engineering\\
Bar-Ilan University\\
amir.weiss@biu.ac.il
}

\begin{document}
\ninept
\maketitle

\begin{abstract}
The need to digitize signals with intricate spectral characteristics often challenges traditional analog-to-digital converters (ADCs). The recently proposed modulo-ADC architecture offers a promising alternative by leveraging inherent features of the input signals. This approach can dramatically reduce the number of bits required for the conversion while maintaining the desired fidelity. However, the core algorithm of this architecture, which utilizes a prediction filter, functions properly only when the respective prediction error is bounded. In practice, this assumption may not always hold, leading to considerable instability and performance degradation. To address this limitation, we propose an enhanced modulo-unfolding solution \emph{without} this assumption. We develop a reliable detector to successfully unfold the signals, yielding a robust solution. Consequently, the reinforced system maintains proper operation in scenarios where the original approach fails, while also reducing the quantization noise. We present simulation results that demonstrate the superior performance of our approach in a representative setting.
\end{abstract}
\vspace{-0.1cm}

\section{Introduction}\label{sec:intro}
\vspace{-0.1cm}
In a growing variety of modern systems, signals exhibit intricate spectral characteristics \cite{wang2010advances,li2019integrated,salomon2021improved,liu2022integrated}. For example, in communication and sensing systems \cite{zheng2019radar,zhang2018wideband,pegoraro2022sparcs}, signals often occupy wide but sparsely used bandwidths, posing challenges for efficient digitization at the receivers \cite{abari2013analog}. In such cases, the resolution requirements of traditional analog-to-digital converters (ADCs) can become prohibitively high, making them inefficient with respect to the signal acquisition fidelity requirements. Consequently, this challenge has prompted a rethinking of traditional ADCs.

The recently emerged concept of a \emph{modulo} ADC \cite{ordentlich2018modulo,shah2019signal,lu2020high} (also termed ``unlimited sampling" \cite{bhandari2017unlimited,bhandari2018unlimited,bhandari2020unlimited}) has been one of the leading efforts on this front. By ``folding" the input signal modulo a fixed range, and further exploiting temporal structures of the unfolded signal \cite{romanov2019above}, these architectures can, in principle, significantly reduce the number of bits required for the conversion, while still retaining the necessary fidelity.

More recently, the informed (or oracle) modulo ADC architecture \cite{ordentlich2018modulo}, which relies on prior knowledge of the input signal's second-order statistics, has been extended to a \emph{blind} scenario \cite{weiss2021blind}, wherein such prior knowledge is assumed to be unknown. This architecture, referred to as blind modulo ADC, employs adaptive algorithms to iteratively learn, in real time, the underlying necessary statistical features of the input signal, which are then used in the conversion. Notably, the blind and more robust architecture achieves \emph{the same} performance---in terms of the rate--minimum mean-square error (MSE) distortion trade-off---as the informed architecture. Still, both architectures are, by design, vulnerable to overloads (explained in detail in Section \ref{subsec:stateoftheart}), which destabilize the adaptive process, leading to large quantization noise and undermining the overall system performance.

In this paper, our main contribution is a generalized, robust algorithm for the blind modulo-ADC architecture that addresses this limitation. By relaxing key assumptions related to the failure mechanism in \cite{weiss2021blind}, and developing a statistically enhanced unfolding technique for the modulo-quantized samples, our solution improves the stability of the system, even during overload events. By further reducing the quantization noise, our approach yields a higher-resolution digitized signal, allowing for more efficient ADC in challenging environments. We present the theoretical foundation of our approach, analytically develop the enhanced unfolding algorithm, and present simulation results demonstrating substantial improvements over the originally proposed solution \cite{weiss2021blind}.
\vspace{-0.15cm}
\section{The Blind Modulo ADC Problem}\label{sec:problem}
\vspace{-0.15cm}
We now formulate the problem of ADC using modulo compression. For this, let us first define a modulo-ADC and review the underlying concept.\vspace{-0.3cm}
\subsection{Preliminaries on Modulo ADCs}\label{preliminaries}
\begin{figure}[t]
    \centering	
     \includegraphics[width=0.35\textwidth]{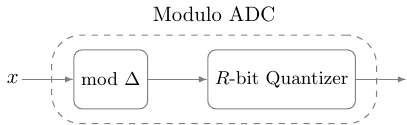}
     \vspace{-0.25cm}
	\caption{A schematic block diagram illustration of the mod-ADC.}
	\label{fig:ModADC_diag_block}\vspace{-0.45cm}
\end{figure}

\vspace{-0.15cm}For a positive number $\Delta\in\positivereals$, we define,
\begin{equation*}
[x]\;{\rm{mod}}\;\Delta\triangleq x-\Delta\cdot\left\lfloor \frac{x}{\Delta}\right\rfloor\in[0,\Delta), \quad \forall x\in\reals,
\end{equation*}
as the $[\cdot]\;{\rm{mod}}\;\Delta$ operator, where $\left\lfloor x\right\rfloor$ is the floor operation, which returns the largest integer smaller than or equal to $x$. An $R$-bit modulo ADC with resolution parameter $\alpha$, termed $(R,\alpha)$ mod-ADC, computes
\begin{equation}\label{modADCdefinition}
[x]_{R,\alpha}\triangleq\left[\left\lfloor \alpha x\right\rfloor\right]\;{\rm{mod}}\;2^R\in\{0,1,\ldots,2^R-1\},
\end{equation}
and produces the binary representation of \eqref{modADCdefinition} as its output, see Fig.~\ref{fig:ModADC_diag_block}.

An $(R,\alpha)$ mod-ADC can be modeled, using subtractive dithers \cite{lipshitz1992quantization}, as a stochastic channel, whose output $\rndy$ for an input $x$ is given by
\begin{equation}\label{modulorandomchannel}
\rndy=\left[\alpha x+\rndz\right]\;{\rm{mod}}\;2^R,
\end{equation}
where $\rndz\sim{\rm{Unif}}\left((-1,0]\right)$. Since the modulo operation is a form of \emph{lossy} compression, it is generally impossible to recover the ``unfolded" signal $(\alpha x+\rndz)$ from its folded version $\rndy$ \eqref{modulorandomchannel}. However, under reasonably mild conditions, when the input signal exhibits certain properties, e.g., a temporally correlated random process \cite{ordentlich2018modulo} or a deterministic bandlimited signal \cite{bhandari2017unlimited,romanov2019above}, it is possible to \emph{perfectly} recover the unfolded signal\footnote{With high probability for random signals, and to an arbitrary precision for deterministic bandlimited signals.} from its past samples and its current folded sample via causal processing \cite{ordentlich2016integer}.

Furthermore, it was recently shown in \cite{weiss2021blind}, via a constructive solution in the form of an online-learning adaptive algorithm, that the temporal correlation structure of the input signal need only to exist and can otherwise be unknown. Since the correlation function of the input is assumed (more realistically) to be unknown, the architecture in \cite{weiss2021blind} is referred to as \emph{blind} modulo-ADC. An extension of this work for random vector processes, more suitable for sensor arrays, was developed in \cite{weiss2022blind}.\vspace{-0.3cm}

\subsection{Problem Formulation}\label{subsec:problemformulation}
\vspace{-0.1cm}Consider an $(R,\alpha_n)$ mod-ADC as described above, with a fixed modulo range $\Delta=2^R$, but an adaptable, possibly time-varying resolution parameter $\alpha_n\in\positivereals$. The mod-ADC is fed with the input discrete-time signal $\{\rndx_n\triangleq \rndx(nT_s)\in\reals\}_{n\in \naturals}$, acquired by sampling the analog, continuous-time signal $\rndx(t)$ every $T_s$ seconds. We assume that $\rndx_n$ is a zero-mean stationary stochastic process with an unknown autocorrelation function $R_{\rndx}[\ell]\triangleq\Exop\left[\rndx_n\rndx_{n-\ell}\right]$. The observed, distorted signal at the output of the mod-ADC reads
\begin{equation}\label{outputofmodADC}
\rndy_n=[\alpha_n \rndx_n+\rndz_n]\;{\rm{mod}}\;2^R, \; \forall n\in\naturals,
\end{equation}
where the quantization noise $\rndz_n\hspace{-0.05cm}\sim\hspace{-0.05cm}{\rm{Unif}}((-1,0])$ \cite{gray1998quantization} is independent, identically distributed (iid). We also define the unfolded quantized signal,
\begin{equation}\label{defofv}
\rndv_n\triangleq\alpha_n \rndx_n + \rndz_n, \; \forall n\in\naturals.
\end{equation}
Observe that, in general, $\rndv_n$ is not stationary. However, when $\alpha_n$ is held fixed on a specific time interval, then $\rndv_n$ can be regarded as stationary on that particular interval, a key observation for the solution to be presented.

Our goal is to estimate the input signal $\{\rndx_n\}$ as accurately as possible based on the observed sequence $\{\rndy_n\}$ at the output of the mod-ADC using causal processing. The problem, then, is stated as follows.

\tcbset{colframe=gray!90!blue,size=small,width=0.49\textwidth,halign=flush center,arc=2mm,outer arc=1mm}
\begin{tcolorbox}[upperbox=visible,colback=white,halign=justify]
\noindent\textbf{Problem Statement:} {For a fixed number of bits $R$, design an adaptive, causal estimator of $\{\rndx_n\}$ based on the mod-ADC output $\{\rndy_n\}$ with the lowest possible MSE, without prior knowledge of $R_x[\ell]$.}
\end{tcolorbox}

Note that since $\rndv_n$ is simply a scaled version of $\rndx_n$ contaminated by white noise \eqref{defofv}, the problem essentially boils down to recovering $\rndv_n$, i.e., unfolding $\rndy_n$. The input signal $\rndx_n$ is then readily estimated as $\widehat{\rndx}_n\triangleq \left(\widehat{\rndv}_n+\tfrac{1}{2}\right)/\alpha_n$, 
where $\alpha_n$ is a known system parameter, and $\frac{1}{2}$ is to compensate for the quantization noise (non-zero) expectation $\Exop[\rndz_n]=-\frac{1}{2}$.
\vspace{-0.5cm}
\section{An Enhanced Algorithmic Solution}\label{sec:enhanced}
\vspace{-0.15cm}
In this section, we present our reinforced solution for a robust blind mod-ADC. We start by providing a bird's-eye view of the state of the art \cite{weiss2021blind}, discussing its main failure mechanism, and pinpointing the underlying (simplistic) assumption that leads to its performance limitation. Building on this critical observation, we then develop our proposed solution.\vspace{-0.25cm}

\subsection{Prior Art and Its Failure Mechanism}\label{subsec:stateoftheart}
{
\renewcommand{\algorithmcfname}{Pseudo Algorithm} 
\setlength{\textfloatsep}{0pt}
\begin{algorithm}[t]
	\KwIn{Signal: $\{\left.\rndx(t)\right|_{t=nT_s}\}$, System Parameters (e.g., $R, \alpha_0$)}
	\KwOut{$\{\widehat{\rndx}_{n}\}$}
	\nl Initialization; \algorithmiccomment{e.g., $\alpha_n=\alpha_0$}\\
    \For{$n=1,2,\ldots$}
    {
    \nl Blind modulo unfolding; \algorithmiccomment{See Algorithm \ref{Algorithm2}}\\
    \If{\emph{$\widehat{\vech}_{n,\text{LMS}}$ converged}} 
    {
	\nl Update resolution parameter $\alpha_n$; \algorithmiccomment{Increase/decrease}\\
    }
    \nl Update the prediction filter $\widehat{\vech}_{n,\text{LMS}}$ \algorithmiccomment{LMS update \cite{haykin2003least}}\\
	\nl {Error propagation mitigation;} \algorithmiccomment{Overload detection}}
	\caption{Blind Modulo ADC \cite[Alg.~3]{weiss2021blind} \label{algorithm1}\vspace{-0.25cm}}
\end{algorithm}
\setlength{\textfloatsep}{0pt}
}
\begin{algorithm}[t!]
	\nl Compute the linear predictor of $\rndv_n$ based on $\mybar{\vecv}$,
    {\setlength{\belowdisplayskip}{4pt} \setlength{\belowdisplayshortskip}{4pt}
    \setlength{\abovedisplayskip}{4pt} \setlength{\abovedisplayshortskip}{4pt}	
    \begin{equation}\label{linearestimation}
	\widehat{\rndv}_n^p\triangleq\tp{\vech}\mybar{\vecv}-\tfrac{1}{2},
	\end{equation}}\\
	\nl Compute
    {\setlength{\belowdisplayskip}{4pt} \setlength{\belowdisplayshortskip}{4pt}
    \setlength{\abovedisplayskip}{1pt} \setlength{\abovedisplayshortskip}{1pt}	
	\begin{align}
	\rndw_n &\triangleq [\rndy_n-\widehat{\rndv}_n^p]\;{\rm{mod}}\;2^R,\label{eq:defofw}\\
	{\widehat{\rnde}}_n^{p} &\triangleq \left(\left[\rndw_n+\tfrac{1}{2}2^R\right]\;{\rm{mod}}\;2^R\right)-\tfrac{1}{2}2^R;\label{estimatedestimationerrors}
	\end{align}}\\
	\nl Return $\widehat{\rndv}_n\triangleq\widehat{\rndv}_n^p+\widehat{\rnde}_n^{p},\;\widehat{\rndv}_n^p$.
     \caption{{\bf Modulo Unfolding} \newline $\widehat{\rndv}_n,\,\widehat{\rndv}_n^p=\text{ModUnfold}(\rndy_n, \mybar{\vecv}, \vech, R)$ \newline {Oracle: $\mybar{\vecv}\hspace{-0.05cm}=\hspace{-0.05cm}\mybar{\rvecv}_{n-1}, \vech\hspace{-0.05cm}=\hspace{-0.05cm}\vech_{\text{opt}}$, \quad Blind: $\mybar{\vecv}\hspace{-0.05cm}=\hspace{-0.05cm}\widehat{\mybar{\rvecv}}_{n-1}, \vech\hspace{-0.05cm}=\hspace{-0.05cm}\widehat{\rvech}_{n,\text{LMS}}$}\label{Algorithm2}}
\end{algorithm}
\setlength{\textfloatsep}{0pt}
\vspace{-0.1cm}The blind mod-ADC architecture \cite{weiss2021blind} conceptually mimics the operation of the oracle algorithm that has access to $R_{\rndx}[\ell]$ \cite[Section II.A]{ordentlich2018modulo}, and is described in Pseudo Algorithm \ref{algorithm1}, whose interpretation is the following. For a fixed resolution parameter $\alpha_n$, given that at any time instance $n$ the unfolded signal $\rndv_n$ can be \emph{exactly} recovered, we have access to the prediction error $\rnde_n^p\triangleq\rndv_n-\widehat{\rndv}_n^p$, where $\widehat{\rndv}_n^p$ \eqref{linearestimation} is a linear predictor of $\rndv_n$ based on the last consecutive $p$ samples $\widehat{\mybar{\rvecv}}_{n-1}\triangleq\tp{\left[\widehat{\rndx}_{n-1} \cdots \widehat{\rndx}_{n-p}\right]}\in\reals^{p\times 1}$.\footnote{While the notation $\widehat{\mybar{\rvecv}}_{n-1}$ for the vector $\tp{[\widehat{\rndx}_{n-1},\ldots,\widehat{\rndx}_{n-p}]}$ may be confusing without reasoning, we follow the original notation for consistency with \cite{weiss2021blind}, and refer the interested reader to \cite{weiss2021blind} for the rationale behind this choice.} Thus, we may learn the corresponding \emph{optimal} linear minimum MSE length-$p$ finite impulse response filter $\vech_{\text{opt}}\in\reals^{p\times 1}$ online, e.g., via the least mean squares (LMS) algorithm \cite{haykin2003least}. Upon convergence of the LMS \cite{feuer1985convergence} to the optimal prediction filter, the resolution parameter $\alpha_n$ can be slightly increased (so as to decrease the quantization noise), and as long as the magnitude of the prediction error $|\rnde_n^p|$ is sufficiently small, $\rndv_n$ could still be recovered using Algorithm \ref{Algorithm2}. To see this, we recall from \eqref{outputofmodADC} and \eqref{defofv} that
\begin{equation}
    \rndy_n=[\rndv_n]\;{\rm{mod}}\;\Delta = [\widehat{\rndv}^p_n + \rnde_n^p]\;{\rm{mod}}\;\Delta,
\end{equation}
since $\rndv_n=\widehat{\rndv}^p_n + \rnde_n^p$ by definition. Therefore, since for any $a,b\in\reals$ and $\Delta>0$,
\begin{equation}
    [[a]\;{\rm{mod}}\;\Delta + b]\;{\rm{mod}}\;\Delta = [a+b]\;{\rm{mod}}\;\Delta,
\end{equation}
it follows that, if $|\rnde_n^p|<\frac{1}{2}\Delta$, we have,
\begin{align}
    &[\rndy_n-\widehat{\rndv}_n^p]\;{\rm{mod}}\;\Delta = [\rnde_n^p]\;{\rm{mod}}\;\Delta\label{eq:moduloerror}\\
    \Longrightarrow\;\;\; &{\widehat{\rnde}}_n^{p} = \left(\left[[\rnde_n^p]\;{\rm{mod}}\;\Delta+\tfrac{1}{2}\Delta\right]\;{\rm{mod}}\;\Delta\right)-\tfrac{1}{2}\Delta = \rnde_n^p. \label{eq:whendoespredictionerrorisrecovered}
\end{align}

At this point, fixing $\alpha_n$ again to its new value, the (now slightly different) optimal prediction filter can be learned. This process is repeated until the asymptotic, steady-state resolution is attained \cite[Eq.~(37)]{weiss2021blind}. Due to space limitation, here only the general structure of this adaptive mechanism is described in Pseudo Algorithm \ref{algorithm1}. For a detailed description, which is not critical for what follows, see \cite[Section IV.E]{weiss2021blind}.

A natural question to ask is the following: \emph{When does the algorithmic architecture described above fail?} Moreover, perhaps an equally important question to ask is \emph{Why does it fail?} A closer examination of this mechanism, and specifically of the core component Algorithm \ref{Algorithm2}, reveals its vulnerability in step \eqref{estimatedestimationerrors}. Indeed, as explained above, the equality \eqref{eq:whendoespredictionerrorisrecovered} holds only when $|\rnde_n^p|<\frac{1}{2}\Delta$. For this reason, the ``overload event" is defined as
\begin{equation}\label{eq:overloadeventdefinition}
\setE_{\tiny {\rm{OL}}_n}\triangleq\left\{\left|\rnde_{n}^p\right|\geq\tfrac{1}{2}\Delta\right\}=\left\{\widehat{\rnde}_{n}^{p} \neq \rnde_{n}^p \right\}.
\end{equation}

In order to control this failure mechanism, we set $\Delta/2=\kappa\cdot\sigma_{p,n}$, where $\sigma_{p,n}$ is the standard deviation of the prediction error $\rnde_n^p$, which can be reliably estimated online, and where $\kappa$ is a confidence (system) parameter that determines the overload probability ($\uparrow\kappa\;\Rightarrow\;\downarrow\prob{\setE_{\tiny {\rm{OL}}_n}}$). Note the inherent trade-off in choosing $\kappa$: On one hand, a larger value gives higher stability (lower overload probability), but degenerates the modulo operation, thus leading to a higher MSE. On the other hand, a smaller value gives lower MSE conditioned on no overloads, but at the cost of frequent overload events, namely higher instability, which eventually leads to large errors and a higher MSE. Indeed, as shown in \cite[Eq.~(39)]{weiss2021blind}, the asymptotic resolution of the blind mod-ADC reads,
{\setlength{\belowdisplayskip}{4pt} \setlength{\belowdisplayshortskip}{4pt}
 \setlength{\abovedisplayskip}{4pt} \setlength{\abovedisplayshortskip}{4pt}	
\begin{equation}\label{eq:asymptoticresolution}
    \alpha_{\infty} = \frac{1}{\kappa}\cdot\left(\frac{1}{2}2^R\right)\cdot\frac{1}{\mybar{\sigma}_{p,\infty}}
\end{equation}}
\hspace{-0.1cm}where $\mybar{\sigma}^2_{p,\infty}$ denotes the asymptotic variance of the optimal linear $p$-length predictor of $\mybar{\rndv}_n\triangleq \frac{\rndv_n+\frac{1}{2}}{\alpha_n}$. The trade-off is well-reflected in \eqref{eq:asymptoticresolution}.

While the interplay in \eqref{eq:asymptoticresolution} seems reasonable, it should be emphasized that it is a trade-off of the specific solution developed in \cite{weiss2021blind}. In general, however, there is no reason to think that it is fundamental to the problem itself (regardless of any particular solution), and that a better trade-off cannot be attained. In fact, in this paper, we argue and demonstrate that in certain, though common settings, we can increase the asymptotic resolution. In other words, it is possible to set a smaller $\kappa$ and further reduce the quantization noise in $\widehat{\rndx}_n$, while retaining stability of the system.

Revisiting \eqref{estimatedestimationerrors}, given that an overload \eqref{eq:overloadeventdefinition} has occurred, we indeed have $\widehat{\rnde}_{n}^{p} \neq \rnde_{n}^p$. However, this inequality is not general and unstructured in the sense that $\widehat{\rnde}_{n}^{p}$ can take any value different from $\rnde_{n}^{p}$. Rather, it must that $\widehat{\rnde}_{n}^{p}=\rnde_{n}^p + \rndm_n\Delta$, for some $\rndm_n\in\integers\backslash\{0\}$. In fact, more generally,
\begin{equation}\label{eq:generalformofestimatedpredictionerror}
     \widehat{\rnde}_{n}^{p}=\rnde_{n}^p + \rndm_n\Delta, \; \rndm_n\in\integers \; \Longrightarrow \; \begin{cases}
     \rndm_n=0 \iff \setE_{\tiny \mybar{{\rm{OL}}}_n}\\
     \rndm_n\neq0 \iff \setE_{\tiny {\rm{OL}}_n}\\
     \end{cases},
\end{equation}
where $\setE_{\tiny \mybar{{\rm{OL}}}_n}$ denotes the no overload event (the compliment of $\setE_{\tiny {\rm{OL}}_n}$). This key observation is the premise of our proposed solution. It shows, first, that Algorithm \ref{Algorithm2}---and step \eqref{estimatedestimationerrors} in particular---implicitly assume that $\rndm_n=0$ at all times. For this very reason, with the original solution \cite{weiss2021blind}, an overload automatically implies an error in the recovery of $\rndv_n$, inflicting a large instantaneous squared quantization error $(\widehat{\rndx}_n-\rndx_n)^2$. In addition, this observation suggests that if $\rndm_n$ can be reliably detected even at instances when $\rndm_n\neq0$, then $\rndv_n$ can \emph{still} be exactly recovered. In such cases, a high-resolution quantized version of $\rndx_n$ is maintained via
\begin{align}
    &\widehat{\rndv}_n(\widehat{\rndm}_n)\triangleq\widehat{\rndv}_n^p+\underbrace{\widehat{\rnde}_n^{p}-\widehat{\rndm}_n\Delta}_{=\rnde_n^p\mid\widehat{\rndm}_n=\rndm_n}=\rndv_n \\
    &\quad\quad\underset{\widehat{\rndv}_n(\widehat{\rndm}_n)=\rndv_n}{\Longrightarrow} \; \widehat{\rndx}_n = \frac{\widehat{\rndv}_n+\tfrac{1}{2}}{\alpha_n} = \frac{\rndv_n+\tfrac{1}{2}}{\alpha_n} = \rndx_n + \underbrace{\frac{\rndz_n+\tfrac{1}{2}}{\alpha_n}}_{\triangleq\rndq_n},
\end{align}
where $\widehat{\rndm}_n$ denotes a detector of $\rndm_n$, and $\rndq_n$ is the quantization noise conditioned on $\widehat{\rndv}_n(\widehat{\rndm}_n)=\rndv_n$, with $\Exop[\rndq_n|\widehat{\rndv}_n(\widehat{\rndm}_n)=\rndv_n]=0$ and $\Varop(\rndq_n|\widehat{\rndv}_n(\widehat{\rndm}_n)=\rndv_n) = \frac{1}{12\alpha_n^2}$. The essence of the above is as follows: \emph{an overload does not necessarily imply that $\rndv_n$ cannot be exactly recovered}:
\begin{equation}\label{eq:oldoesnotimplyinequality}
\setE_{\tiny {\rm{OL}}_n} \not\Rightarrow \left\{ \widehat{\rndv}_n\neq\rndv_n \right\}.
\end{equation}

\subsection{Proposed Solution: Approximate MAP Unfolding}\label{subsec:approximatemapunfolding}
The relations \eqref{eq:generalformofestimatedpredictionerror} and \eqref{eq:oldoesnotimplyinequality} are the key findings from which our proposed solution naturally stems. Specifically, we now develop a detector $\widehat{\rndm}_n$ for $\rndm_n$, with which we define new (improved) estimators for $\rnde^p_n$ and $\rndv_n$.

We start by observing that, given $\{\rndy_n,\mybar{\rvecv}_{n-1}\triangleq\tp{[\mybar{\rndv}_{n-p} \cdots \mybar{\rndv}_{n-1}]}\}$, the optimal detector of $\rndm_n$ in the sense of minimum error probability, namely the maximum \textit{a posteriori} (MAP) \cite{van2004detection} detector, is given by,
\begin{align}
\widehat{\rndm}_{n,{\text{\tiny MAP}}} &\triangleq \arg \max_{m\in\integers} \; \prob{\rndm_n=m|\rndy_n,\mybar{\rvecv}_{n-1}}\label{eq:oraclemapdef}\\
&=\arg \max_{m\in\integers} \; \prob{\mybar{\rvecv}_{n-1,n}|\rndm_n=m}\prob{\rndm_n=m}\label{eq:transitionfromytov},
\end{align}
where we have defined
\begin{align}
\mybar{\rvecv}_{n-1,n}&\triangleq\tp{[\tp{\mybar{\rvecv}_{n-1}} \; \widehat{\mybar{\rndv}}_n]}\in\reals^{(p+1)\times1},\\
\widehat{\mybar{\rndv}}_n&\triangleq \tfrac{\widehat{\rndv}_n^p + \widehat{\rnde}_n^{p}-\tfrac{1}{2}}{\alpha_n} = \tfrac{\widehat{\rndv}_n^p + \rnde_n^{p}-\rndm_n\Delta-\tfrac{1}{2}}{\alpha_n},\label{eq:relationforconditionaldistribution}
\end{align}
and \eqref{eq:transitionfromytov} is due to fact that $\widehat{\mybar{\rndv}}_n$ is a deterministic function of $\rndy_n$ and $\widehat{\rndv}_n^p$, and the latter is a deterministic function of $\mybar{\rvecv}_{n-1}$ (via \eqref{linearestimation}).

At this point, we note that in the blind scenario under consideration, where the distribution of $\rndx_n$---and therefore of $\rndv_n$---is unknown, it is generally unclear how to proceed to develop a detector for $\rndm_n$. Thus, to this end, we henceforth assume that $\{\rndv_n\}$ is a Gaussian process.\footnote{While for different inputs the distribution of $\rndv_n$ may be different from Gaussian, this assumption would lead to a robust solution.} Now, since $\{\rndv_n\}$ is Gaussian, $\{\rnde_n^p\}$ is also Gaussian. Therefore, $\rndm_n$ is a discrete random variable whose support is $\integers$, with a probability mass function\footnote{$Q(\cdot)$ denotes the standard $Q$-function \cite{karagiannidis2007improved}.}
\begin{align}
\prob{\rndm_n=m} &= \prob{\rnde_n^{p}\in\left(m\Delta-\tfrac{\Delta}{2},m\Delta+\tfrac{\Delta}{2}\right)} \label{eq:usemodulorelationfortransition}\\
&= \left|Q\left(\tfrac{(m-0.5)\Delta}{\sigma_{p,n}}\right)-Q\left(\tfrac{(m+0.5)\Delta}{\sigma_{p,n}}\right)\right|\triangleq e^{-\frac{F(m)}{2}},\label{eq:definitionofF}
\end{align}
where we have used $\rnde_n^p=\widehat{\rnde}_n^p-\rndm_n\Delta$ of \eqref{eq:generalformofestimatedpredictionerror} in \eqref{eq:usemodulorelationfortransition}. Furthermore, since $\{\rndv_n\}$ is Gaussian, it follows that $\widehat{\mybar{\rndv}}_n|\rndm_n\sim\normal(\rndm_n\Delta,R_{\rndx}[0]+\frac{1}{12\alpha_n^2})$ from \eqref{eq:relationforconditionaldistribution}. Hence, $\mybar{\rvecv}_{n-1,n}|\rndm_n\sim\normal\left(\vecmu_{\bar{\rndv}}(\rndm_n),\matC_{\bar{\rndv}_n}\right)$, where
\begin{align}
&\vecmu_{\bar{\rndv}}(\rndm_n)\triangleq\Exop\left[\mybar{\rvecv}_{n-1,n}|\rndm_n\right] = \tp{[\tp{\veczero_p} \; \rndm_n\Delta]}\in\reals^{(p+1)\times 1},\\
&\matC_{\bar{\rndv}_n}\triangleq\Exop\left[(\mybar{\rvecv}_{n-1,n}-\vecmu_{\bar{\rndv}}(\rndm_n))\tp{(\mybar{\rvecv}_{n-1,n}-\vecmu_{\bar{\rndv}}(\rndm_n))}|\rndm_n\right],
\end{align}
and we note that $\matC_{\bar{\rndv}_n}\in\reals^{(p+1)\times (p+1)}$ is constant with respect to $\rndm_n$, since $\left[\matC_{\bar{\rndv}_n}\right]_{ij}=R_{\rndx}[i-j]+\mathbbm{1}_{i=j}\cdot\frac{1}{12\alpha^2_{n-p+i-1}}$ for $1\leq i,j\leq p+1$.\footnote{We use $\mathbbm{1}_{[\cdot]}$ to denote the standard indicator function.}

Having obtained closed-form expressions for the prior distribution $\prob{\rndm_n=m}$ and the likelihood $\prob{\mybar{\rvecv}_{n-1,n}|\rndm_n=m}$, we can now explicitly express the MAP detector \eqref{eq:transitionfromytov}. By denoting $\mybar{\rvecv}_{n-1,n}(\rndm_n)\triangleq\mybar{\rvecv}_{n-1,n}-\vecmu_{\bar{\rndv}}(\rndm_n)$ for brevity, taking the (natural) logarithm of \eqref{eq:transitionfromytov}, and omitting irrelevant constants with respect to $\rndm_n$, we obtain,
\begin{align}
\widehat{\rndm}_{n,{\text{\tiny MAP}}}= \arg \min_{m\in\integers} \; \tp{\mybar{\rvecv}_{n-1,n}(m)}\inv{\matC_{\bar{\rndv}_n}}\mybar{\rvecv}_{n-1,n}(m) + F(m).
\end{align}
We now recall that, in practice, we have access to neither the unobservable data $\mybar{\rvecv}_{n-1}$ (the first $p$ entries of $\mybar{\rvecv}_{n-1,n}(m)$), the unknown covariance matrix $\matC_{\bar{\rndv}_n}$, which depends on the known resolution parameters $\{\alpha_n\}_{i=n-p}^N$ but also on the unknown correlation function $\{R_{\rndx}[\ell]\}$, nor $\sigma_{p,n}$, which is required for $F(m)$, defined in \eqref{eq:definitionofF}. Therefore, $\widehat{\rndm}_{n,{\text{\tiny MAP}}}$ is in fact an oracle detector, which is not realizable. 

Nonetheless, the blind mod-ADC adaptive mechanism is initialized and designed in such a way that $\widehat{\mybar{\rvecv}}_{n-1}=\mybar{\rvecv}_{n-1}$ with high probability, at least before reaching to steady state (see \cite[Section IV]{weiss2021blind}). Thus, not only $\widehat{\mybar{\rvecv}}_{n-1}$ can be used instead of $\mybar{\rvecv}_{n-1}$, one may also replace $\matC_{\bar{\rndv}_n}$ with
{\setlength{\belowdisplayskip}{4pt} \setlength{\belowdisplayshortskip}{4pt}
 \setlength{\abovedisplayskip}{4pt} \setlength{\abovedisplayshortskip}{4pt}	
\begin{equation}
    \widehat{\matC}_{\bar{\rndv}_n}\triangleq\frac{1}{n-p-2}\sum_{i=p+2}^{n-1}\widehat{\mybar{\rvecv}}_{i-1,i}\tp{\widehat{\mybar{\rvecv}}_{i-1,i}}, \; n\geq p+3,\label{eq:defofestimatedcovariance}
\end{equation}}
\hspace{-0.1cm}which can be (consistently) estimated iteratively online, and similarly replace $\sigma_{p,n}$ with $\widehat{\sigma}_{p,n}$ (see \cite[Eq.~(22)]{weiss2021blind}). Based on these observable quantities, we propose the approximate MAP (AMAP) detector,
\begin{align}\label{eq:defofamap}
\hspace{-0.05cm}\widehat{\rndm}_{n,{\text{\tiny AMAP}}}= \arg \min_{m\in\setM} \; \tp{\widehat{\mybar{\rvecv}}_{n-1,n}(m)}\inv{\widehat{\matC}_{\bar{\rndv}_n}}\widehat{\mybar{\rvecv}}_{n-1,n}(m) + \widehat{F}(m),
\end{align}
where we have defined $\setM\triangleq\{m\in\integers\mid |m|\leq M\in\naturals\}\subset\integers$ and
\begin{align}
\widehat{\mybar{\rvecv}}_{n-1,n}&\triangleq\tp{[\tp{\widehat{\mybar{\rvecv}}_{n-1}} \; \widehat{\mybar{\rndv}}_n]}\in\reals^{(p+1)\times1},\\
\widehat{F}(m)&\triangleq -2\log\left(\left|Q\left(\tfrac{(m-0.5)\Delta}{\widehat{\sigma}_{p,n}}\right)-Q\left(\tfrac{(m+0.5)\Delta}{\widehat{\sigma}_{p,n}}\right)\right|\right).\label{eq:definitionofestimatedF}
\end{align}
Clearly, \eqref{eq:defofamap} is a function of observed data, and is thus realizable. Using this detector, we propose our robust blind unfolding in Algorithm \ref{Algorithm3}.

For a rigor and unbiased presentation of our detector, let us highlight the differences between the oracle detector \eqref{eq:oraclemapdef} and our approximate one \eqref{eq:defofamap}. Clearly, the oracle MAP \eqref{eq:oraclemapdef} is optimal, i.e., it guarantees the minimum error probability in the detection of $\rndm_n$. For this reason, an oracle mod-ADC architecture equipped with \eqref{eq:transitionfromytov} delivers at least the same performance as (or even better than) the performance of the original blind mod-ADC architecture \cite{weiss2021blind}, which is essentially (and implicitly) equipped with a degenerate detector that always detects $\rndm_n$ as $0$. In contrast, \eqref{eq:defofamap} is suboptimal, and is approximate for the following reasons:
\begin{enumerate}
    \setlength{\itemsep}{0.25pt}  
    \setlength{\parskip}{0.25pt}  
    \item Even if the input signal $\{\rndx_n\}$ is Gaussian, strictly speaking, $\{\rndv_n\}$ is generally not Gaussian, since $\{\rndz_n\}$ is uniform;
    \item Unknown quantities are replaced by their estimates (e.g., $\mybar{\rvecv}_{n-1}$, $\sigma_{p,n}$ are replaced by $\widehat{\mybar{\rvecv}}_{n-1},\widehat{\sigma}_{p,n}$, respectively); and
    \item The infinite optimization domain $\integers$ is reduced to the finite $\setM$.
\end{enumerate}
However, these approximations are well-justified, respectively, by:
\begin{enumerate}
    \setlength{\itemsep}{0.25pt}  
    \setlength{\parskip}{0.25pt}  
    \item Near steady state, for $\alpha_n$ sufficiently large, the quantization noise has a negligible effect on the distribution of $\rndv_n$ relative to $\rndx_n$ \eqref{defofv};
    \item Near steady state (and typically well before), $\widehat{\matC}_{\bar{\rndv}_n},\widehat{\sigma}_{p,n}$ are already close to the true values of their respective estimands ; and
    \item $\prob{\rndm_n=m}$ decays exponentially fast with $m$, so beyond a moderate value of $M$, there is essentially no difference whatsoever in any practical regime of operation when using $\setM$ instead of $\integers$.
\end{enumerate}

\begin{algorithm}[t!]    
	\nl Compute $\widehat{\rndv}_n^p$, the linear predictor of $\rndv_n$ as in \eqref{linearestimation};\\
	\nl Compute $\rndw_n$ and ${\widehat{\rnde}}_n^{p}$ as in \eqref{eq:defofw} and \eqref{estimatedestimationerrors}, respectively;\\
    \nl Compute the detector $\widehat{\rndm}_{n,{\text{\tiny AMAP}}}$ via \eqref{eq:defofamap};\\
	\nl Return $\widehat{\rndv}_n\triangleq\widehat{\rndv}_n^p+\widehat{\rnde}_n^{p}-\widehat{\rndm}_{n,{\text{\tiny AMAP}}}\Delta,\;\widehat{\rndv}_n^p$.
     \caption{{\bf Robust Blind Modulo Unfolding} \newline $\widehat{\rndv}_n,\,\widehat{\rndv}_n^p=\text{RobustModUnfold}(\rndy_n, \widehat{\mybar{\rvecv}}_{n-1}, \widehat{\rvech}_{n,\text{LMS}}, R, \widehat{\matC}_{\bar{\rndv}_n}, \widehat{\sigma}_{p,n})$\label{Algorithm3}}
\end{algorithm}
\setlength{\textfloatsep}{0pt}

An analytical performance analysis of the proposed scheme is beyond the scope of the present (relatively short) paper, and is deferred to future work. Still, in light of the above, and although we do not claim for optimality of \eqref{eq:defofamap}, we argue that it potentially improves---in some cases, significantly---upon the original architecture \cite{weiss2021blind}, which automatically treats an overload event as an unfolding error, thus inflicting a large quantization noise. We demonstrate this significant improvement next.\vspace{-0.15cm}
\vspace{-0.175cm}
\section{Simulation Results}\label{sec:simulationresults}
\vspace{-0.175cm}
\begin{figure}[t!]
    \centering
    \begin{tikzpicture}[>=latex]
        \node[anchor=south west, inner sep=0] (image) at (0,0) {\includegraphics[width=\columnwidth]{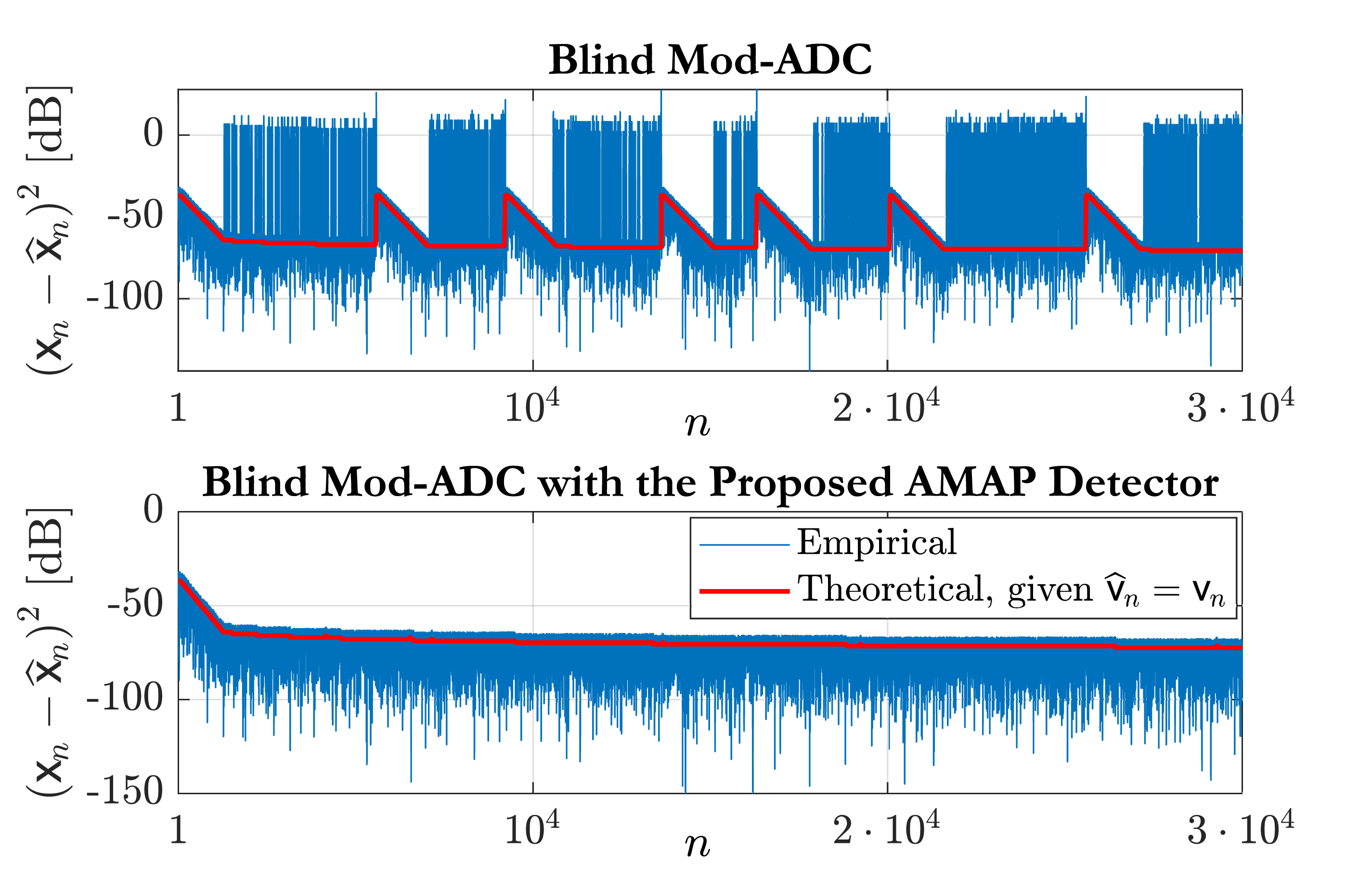}};
        
        \begin{scope}[x={(image.south east)},y={(image.north west)}]
            \draw[orange, thick, ->] (0.175,0.82) -- (0.275,0.75);
            \draw[orange, thick, ->] (0.27,0.82) -- (0.37,0.75);
            \draw[orange, thick, ->] (0.381,0.82) -- (0.481,0.75);
            \draw[orange, thick, ->] (0.45,0.82) -- (0.55,0.75);
            \draw[orange, thick, ->] (0.5475,0.82) -- (0.6475,0.75);
            \draw[orange, thick, ->] (0.58,0.82) -- (0.79,0.75);
            \node[anchor=west] at (0.15,0.835) {\small {\color{orange}\textbf{Error Propagation Mitigation}}};
            \node[anchor=west] at (0.52,0.615) {{\color{Maroon}\tiny$\frac{1}{N}\sum_{n=1}^{N}(\rndx_n-\widehat{\rndx}_n)^2=0.85$ dB}};
            \node[anchor=west] at (0.52,0.14) {{\color{Maroon}\tiny$\frac{1}{N}\sum_{n=1}^{N}(\rndx_n-\widehat{\rndx}_n)^2=-57$ dB}};
        \end{scope}
    \end{tikzpicture}\vspace{-0.35cm}
    \caption{Instantaneous squared quantization error vs.\ discrete-time.}
    \label{fig:se_vs_time}
    \vspace{-0.4cm}
\end{figure}
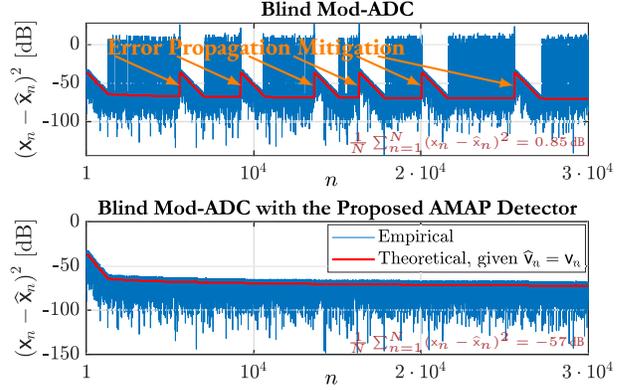
\vspace{-0.05cm}We consider the case of a bandlimited input. Specifically, $\{\rndx_n\}$ is generated by applying a non-ideal, minimum-order bandpass filter with a passband $[\tfrac{\pi}{4},\tfrac{\pi}{2}]\tfrac{{\rm{rad}}}{{\rm{sample}}}$ and a stopband attenuation of $60$ dB, to a unit-variance white Gaussian noise. We then normalize the output, such that $\{\rndx_n\}$ is unit-variance. We set $R=10, \alpha_0=20, p=40, M=2$, and $\kappa=1.5$ such that the overload probability is $\prob{\setE_{\tiny {\rm{OL}}_n}}=2Q(\kappa)=0.1336$ (see \cite[Eq.~(21)]{weiss2021blind}). We deliberately choose such a value of $\kappa$ to challenge the stability of the system with frequent overload events. Other system parameters (e..g., LMS learning rate), which have little effect on our focus, were set according to the guidelines in \cite[Section IV-D]{weiss2021blind}. We note that the results reported below were verified by multiple runs, and were consistently observed for different realizations and other settings.

For an input of length $N=3\cdot10^4$ samples, Fig.~\ref{fig:se_vs_time} presents the evolution in discrete-time of the instantaneous squared quantization error in dB of $\widehat{\rndx}_n$, attained by the blind mod-ADC architecture (top) and our robust enhancement thereof (bottom), equipped with the approximate MAP detector \eqref{eq:defofamap}. The number of overload events was $689$, which gives an empirical probability $\widehat{\mathbb{P}}\left(\setE_{\tiny {\rm{OL}}_n}\right)\approx0.069$, considerably lower from the theoretical one. In this sense, these results are in favor of the original architecture, which has to cope with less overload events. Still, evidently, these are more than enough to drive the system away from stability if it is not equipped with a compensation mechanism, as the one we propose. We also present in red $\Varop(\rndq_n|\widehat{\rndv}_n=\rndv_n)=1/\left(12\alpha_n^2\right)$, the theoretical instantaneous squared quantization error, conditioned on successful unfolding of $\rndy_n$ into $\rndv_n$. As seen, our proposed detector $\widehat{\rndm}_{n,{\text{\tiny AMAP}}}$ exhibits excellent performance, and enables to avoid unnecessary initializations that locally, and significantly, increase the quantization noise. 

\begin{figure}[t!]
    \centering
    \includegraphics[width=\columnwidth]{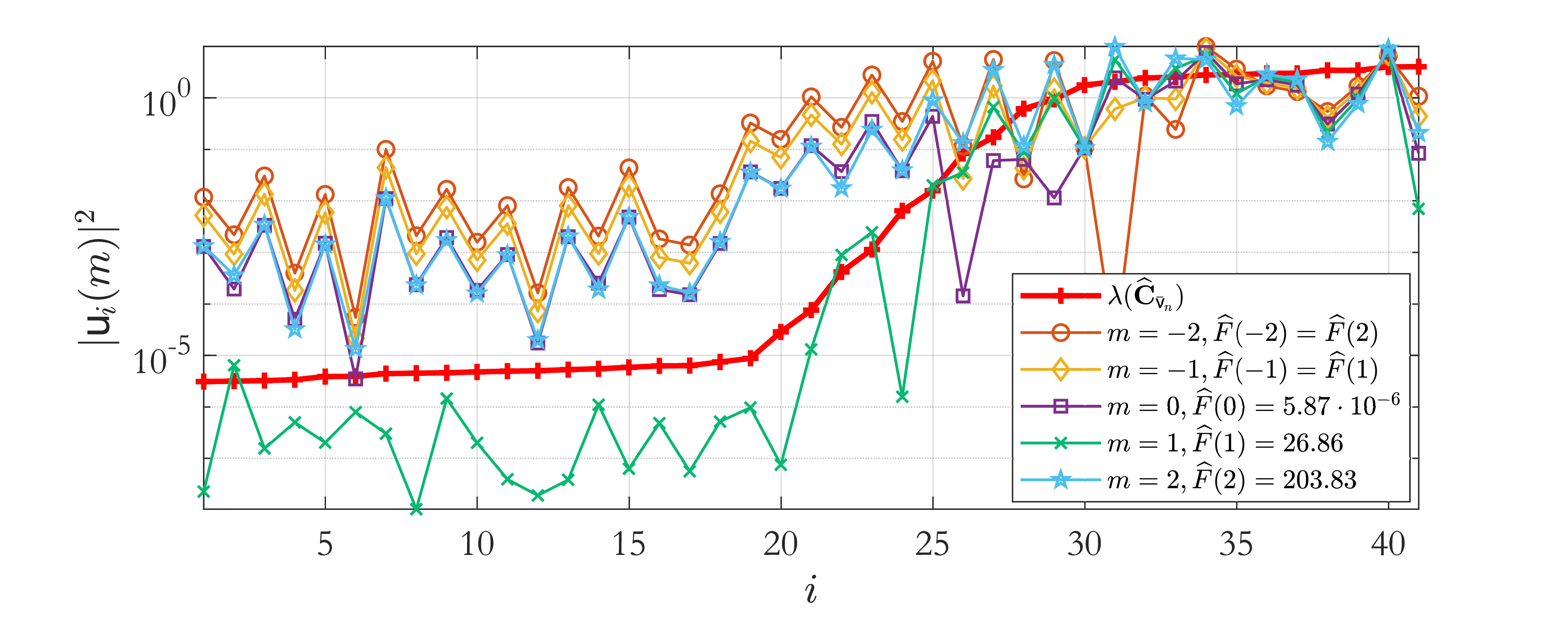}\vspace{-0.35cm}    
    \caption{Eigenvalues of $\widehat{\matC}_{\bar{\rndv}_n}$ (red), and the squared magnitudes of the transformed hypothesized vectors $\{\rvecu(m)=\tp{\matU}\widehat{\mybar{\rvecv}}_{n-1,n}(m)\}_{m\in\setM}$.}
    \label{fig:eigenvalues}
\end{figure}

To explain the operation of \eqref{eq:defofamap}, Fig.~\ref{fig:eigenvalues} presents the spectrum of the estimated covariance \eqref{eq:defofestimatedcovariance}, for which we define the eigendecomposition $\widehat{\matC}_{\bar{\rndv}_n}\triangleq \matU\matLambda\tp{\matU}$, and the squared magnitudes of the hypothesized vectors $\{\rvecu(m)\triangleq\tp{\matU}\widehat{\mybar{\rvecv}}_{n-1,n}(m)\}_{m\in\setM}$. The figure clearly shows that a good fit to the estimated spectral distribution is achieved only for $\widehat{\mybar{\rvecv}}_{n-1,n}(1)$ (via $\rvecu(1)$), whose last entry is correctly modulo-resolved.
\vspace{-0.35cm}
\vspace{-0.025cm}
\section{Conclusion}\label{sec:conclusion}
\vspace{-0.15cm}
\vspace{-0.1cm}We propose an enhanced algorithmic solution for the blind modulo-ADC framework that addresses the instability caused by overload events in existing designs. By careful scrutiny of the failure mechanism, we establish that such overloads must not necessarily lead to large quantization errors, and develop an approximate MAP unfolding technique. As a result, our solution enhances both the robustness and resolution of the system. Simulation results demonstrate that the proposed method significantly reduces the average level of quantization noise while maintaining stability, outperforming the previously proposed architecture. 

A promising direction for future research is to apply the concepts from this work to form a basis for a similar (though probably not identical) enhanced solution to the multidimensional case \cite{romanov2021blind,weiss2022blind,gong2021multi}, e.g., for sensor arrays, where multiple mod-ADCs operate in tandem.

\bibliographystyle{IEEEbib}
\bibliography{refs}

\end{document}